# Cluster Alignments in the Edinburgh/Milano Cluster Redshift Survey

D. R. Martin[1], R. C. Nichol[2], C. A. Collins[3], S. A. Lumsden[4], L. Guzzo[5]
[1] *Department of Physics and Astronomy, Northwestern University, Evanston, Illinois 60208, USA.*
[2] *Department of Astronomy and Astrophysics, University of Chicago, 5640 S. Ellis Ave, Chicago, Illinois 60637, USA.*
[3] *Astrophysics Group, School of Physical and Chemical Sciences, Liverpool-John-Moores University, Byrom Street, Liverpool, L3 3AF, England*
[4] *Anglo-Australian Observatory, PO Box 296, Epping, NSW 2121, Australia.*
[5] *Osservatorio Astronomico di Brera I-22055, Merate, Italy*

5 December 1994

**ABSTRACT**

We present here the results of a statistical search for cluster alignments using the Edinburgh/Milano cluster redshift survey. This survey is a unique cluster database which has been objectively constructed to help minimise the systematic biases associated with previous optical cluster catalogues. We find some evidence for cluster alignments out to spatial separations of $< 10h^{-1}$ Mpc, however, it is not statistically significant. On larger scales, we find no evidence, statistically significant or not, for cluster alignments. These results are in most disagreement with the recent observations of West and Plionis; both of whom see significant cluster alignments out to $\simeq 30h^{-1}$ Mpc and beyond in the Abell & Lick catalogues of clusters. Our findings are consistent with other searches for cluster alignments that do not involve these catalogues.

## 1 INTRODUCTION

Clusters of galaxies are key tracers of the large–scale structure in the universe. One of the most intriguing cluster observations in recent years has been the suggestion that neighbouring clusters are aligned up to a separation of $50h^{-1}$ Mpc*. Such alignments were first reported by Binggeli (1982), who commented that clusters 'strongly tend to point to each other'. He claimed that clusters were elongated towards their *nearest* neighbour within a separation of $30h^{-1}$ Mpc, with a general trend for alignment out as far as $50$–$100h^{-1}$ Mpc. If correct, such observations would have the potential of being an important galaxy formation indicator.

The theoretical implications of cluster alignments have been discussed by several authors. Using N–Body simulations, Dekel, West & Aarseth (1984) showed that the *Binggeli Effect* strongly favoured a scenario of structure evolution which formed the largest structures first, followed by fragmentation into smaller systems like clusters and individual galaxies (a Top-Down scenario). They also showed that a hierarchical scenario (or Bottom-Up) could not easily reproduce cluster alignments and that tidal interactions were insignificant on these large scales. However, more recent N-body simulations of specific models of galaxy formation have showed that the picture is not as clear–cut as portrayed by Dekel *et al.*, but have continued to demonstrate that cluster alignments are a powerful constraint on these models of structure formation (West, Dekel & Oemler 1989 & West, Villumsen & Dekel 1991).

Since Binggeli's original result, many authors have investigated the reality of cluster alignments. For example, both Flin (1987) and Rhee & Katgert (1987) claim to have confirmed Binggeli's result, but at a lower level than he suggested ($\leq 30h^{-1}$ Mpc), while West (1989b) detects cluster alignments, within the same host supercluster, up to separations of $60h^{-1}$ Mpc. On the other hand, several authors have found no statistical evidence for the *Binggeli Effect* (Struble & Peebles 1985, Ulmer, McMillan & Kowalski 1989), or find tentative evidence for cluster alignments only on scales of $< 5h^{-1}$ Mpc (Fong, Stevenson & Shanks 1990). Most recently, Plionis (1994) claims to have observed statistically significant cluster alignments out to a separation of $\simeq 30h^{-1}$ Mpc using the largest database of clusters thus far (277 clusters).

Clearly, even after a decade of study, there is no general consensus within the astronomical literature over the observational reality of cluster alignments. There are apparently valid observations supporting alignments on virtually any scale, ranging from zero out as far as $100h^{-1}$ Mpc. Inevitably, these widely discrepant results undermine the potential of cluster alignments as a serious constraint on models of galaxy formation. These discrepancies are probably the combination of two effects. First, there are obvious problems with using a finite number of galaxies (or binned galaxy counts) to determine the shape and orientation of the underlying total mass distribution of a cluster *i.e.* there are too few galaxies to trace the cluster gravitational potential well. This is illustrated in Figure 1 where we present the position angles for 34 Abell clusters (Abell 1958) as determined independently by both Plionis (1994) and West (1989b). We have only included clusters that have robust position angle measurements as claimed by each author (there are a further 25 clusters in common between the two samples but which have one author claiming the position angle is robust while the other does not). This plot clearly demonstrates the difficulty in determining secure position angles from optical galaxy counts since even for the most robust measurements in the literature display a large scatter about the mean, $\simeq 40°$. We return to this point in Section 3 of this paper.

---

* Throughout, we use $H_o = 100 \, \mathrm{km \, s^{-1} \, Mpc^{-1}}$

Secondly, the vast majority of the above studies have involved the Abell and/or Lick catalogues (Abell 1958, Abell et al. 1989, Shane & Wirtanen 1967) which are known to be plagued by systematic biases (Geller, de Lapparent & Kurtz 1984, de Lapparent, Kurtz & Geller 1986, Sutherland 1988, Lumsden et al. 1992). For example, Lumsden et al. (1992) has shown that the completeness of the Abell catalogue, at all richnesses, maybe as low as 50% which would result in a heterogeneous sample of clusters with little certainty of the true neighbourhood of any cluster within the catalogue. Furthermore, several authors have claimed that the Abell catalogue is plagued by projection effects, where intrinsically poor groups/clusters in the haloes of nearer rich clusters are artificially boosted into the Abell catalogue, thus leading to spurious angular clustering. Such projection effects could bias position angle determinations of both the clusters in question, because the galaxy density in the overlap region of the two clusters will be boosted thus artificially skewing both the position angles towards each other. Finally, the true frequency of phantom clusters within the Abell catalogue remains unclear *i.e.* where galaxies/groups aligned along the line of sight give the impression of a rich cluster in projection. Simulations of this effect claim that as many as 50% of clusters seen in projection are spurious (Frenk et al. 1990). Clearly, the combined effect of such problems hinders statistical studies of large scale coherence of clusters using the Abell catalogue.

In this paper, we present the results of a search for cluster alignments from the Edinburgh/Milano cluster redshift survey. This catalogue is a unique database of clusters since it addresses many of the problems discussed above concerning the Abell catalogue. Our motivation for this work was to quantify the degree of any alignment between clusters and therefore, place this area of study on a stronger observational footing. Although by themselves, cluster alignments may not be as powerful a galaxy formation indicator as one hoped (discussed above), they would provide a further observational constraint on theories of structure formation.

In the following section, we describe the catalogue of clusters used in our search for cluster alignments. In Sections 3 and 4, we present the techniques used to derive the positions angles of our clusters and carry out Monte–Carlo simulations to estimate the error on these. In Section 5, we detail the results of our search for statistical significant cluster alignments. Finally, in Section 6, we discuss our result in the light of previous work.

## 2 THE EDINBURGH/MILANO CLUSTER REDSHIFT SURVEY

The Edinburgh/Milano cluster redshift survey (EM survey) is the most comprehensive optical cluster survey to date and has two major advantages over previous optical cluster catalogues. First, the 2–D cluster candidates were selected automatically using strict objective criteria from the Edinburgh/Durham Southern Galaxy Catalogue (EDSGC), a galaxy catalogue which again was constructed objectively from automated scans of photographic plates. The EDSGC contains over 1.5 million galaxies to a magnitude limit of $b_j = 20.5$ and covers a contiguous area of $1400 \, \text{deg}^2$ at the South Galactic Pole (Collins, Nichol & Lumsden 1992). Furthermore, during the selection of the clusters from the EDSGC, an objective approach was taken to the classification of overlapping clusters thus minimising the problems of projection effects discussed extensively above (see Lumsden et al. 1992). In total, 737 overdensities were selected from the EDSGC and this catalogue of clusters/groups constitutes the Edinburgh/Durham Cluster Catalogue (EDCC). Secondly, most clusters within the EM survey have approximately 10 galaxy redshift measurements thus reducing the problems of phantom clusters and galaxy interlopers. The EM survey in total comprises of secure redshift measurements for $\simeq 100$ of the richest clusters selected from the EDCC. This combined approach effectively eliminates many of the systematic biases that have plagued previous cluster catalogues (see Sutherland 1988, Postman et al. 1985) and, for the first time, allows a true statistical sample of clusters to be selected.

For the work presented here, we used the statistically complete sample of 97 clusters as defined in Nichol et al. (1992). The sample was selected with a smaller than usual Abell radius (Abell 1958) of $1.0 h^{-1}$ Mpc and formed the basis for an investigation of the Spatial Correlation Function of clusters (Nichol et al. 1992). Of this sample, 65 clusters have a redshift determined by us and we have removed 6 of these clusters as possible phantom clusters. A literature redshift was available for a further 20 clusters in the full sample. This therefore, left us with a sample of 91 clusters of which 79 have a measured redshift. All data relating to this sample of clusters, including their coordinates, measured redshifts and position angles used in the work described here, are presented as a whole in Collins et al. (1994). The reader is referred to this paper for a complete discussion of the sample and to obtain the data.

## 3 CLUSTER POSITION ANGLES

An obvious prerequisite to measuring cluster alignments is to calculate the position angles for each cluster in our sample of clusters. With accurate positions and magnitudes for all galaxies within the EDSGC, we have an obvious advantage over previous studies in finding the major axis of any given cluster.

### 3.1 Fitting an Ellipse

We extracted all galaxies from the EDSGC that were within one square degree of the cluster centroid for all clusters in our sample. This area was large enough to provide a good determination of the local background of the cluster without being too large as to include other nearby overdensities. For example, the average angular Abell radius of our clusters was $\sim 0.15$ degrees, while from the EDCC, the surface density of clusters/groups is $\simeq 0.5 \, \mathrm{deg}^{-2}$. These data therefore, formed the basis from which we determined the position angle of our clusters.

We then extracted all galaxies that were within the magnitude range $m_3 < m < m_3 + 2$, where $m_3$ was the background corrected magnitude of the third brightest cluster member (Lumsden et al. 1992). This is the same definition as Abell used for his richness estimate of clusters. These galaxies were then fitted with an ellipse using the iterative process described by Carter & Metcalfe (1980). The advantage of this method is that it is insensitive to a uniformly distributed background of galaxies. In addition, we weighted each galaxy by the inverse of its magnitude since this was empirically found to be better than evenly weighting all the galaxies as Carter & Metcalfe had done.

The iterative process was started with an initial guess of a circle of radius $1.2 h^{-1}$ Mpc on the observed centroid of the cluster. This radius is smaller than originally used by Abell but was larger than the one used to define the cluster (see above). This value was found empirically to be the best for fitting our clusters as it concentrated the fit towards the cluster core without being dominated by a few bright galaxies at the core. This spatial radius was converted at an angular aperture ($\theta_A$) using the standard formula;

$$\theta_A = \tan^{-1}\left(\frac{1.2\, H_o}{c\, z}\right), \qquad (1)$$

were $z$ was the measured redshift of the cluster. The galaxies within this ellipse were then used to compute the major ($a$) and minor ($b$) axes and position angle ($\theta_{PO}$) as defined by:

$$\begin{aligned} a^2 &= 2\,(M_{xx} + M_{yy}) \\ &\quad + 2\left[(M_{xx} - M_{yy})^2 + 4 M_{xy}^2\right]^{\frac{1}{2}}, \\ b^2 &= 2\,(M_{xx} + M_{yy}) \\ &\quad - 2\left[(M_{xx} - M_{yy})^2 + 4 M_{xy}^2\right]^{\frac{1}{2}}, \\ \tan(2\theta_{PO}) &= \frac{2 M_{xy}}{M_{xx} - M_{yy}}, \end{aligned} \qquad (2)$$

where the second moments $M_{xx}$, $M_{yy}$ and $M_{xy}$ are:

$$\begin{aligned} M_{xx} &= W^{-1} \sum_{i=1}^{N} (x_i - \bar{x})^2 w_i^2, \\ M_{yy} &= W^{-1} \sum_{i=1}^{N} (y_i - \bar{y})^2 w_i^2, \\ M_{xy} &= W^{-1} \sum_{i=1}^{N} (x_i - \bar{x})(y_i - \bar{y}) w_i^2, \qquad (3) \\ W &= \sum_{i=1}^{N} w_i. \end{aligned}$$

(4)

In the above formulae, $x_i$ are the Right Ascensions, $y_i$ the Declinations and $w_i$ the weights of the galaxies.

The procedure was continued until the process converged. Between each iteration, we conserved the area within the fitted ellipse by ensuring that $\sqrt{ab} =$ constant. If this step was neglected, then the ellipse tended to shrink to a point at the centre of the cluster. In addition, we also allowed the centroid ($\bar{x}$, $\bar{y}$) to move, allowing the fitting process to find the best solution for the distribution of galaxies in question. We found that in all cases, the centroid did not move by more than 0.1 degrees from the original centroid. The criteria for convergence was set to be that the parameters of the fit had to change by less than 0.1% between iterations. All clusters converged to a stable solution within 10 iterations.

Finally, we counted the number of galaxies within the final fitted ellipse, which was typically 50 galaxies. We then set a cut at 20 galaxies within the final fitted ellipse and discarded any clusters that did not satisfy this criteria. This ensured that there were enough galaxies within the ellipse to confidently determine the clusters eccentricity ($b/a$) and position angle. Only 4 clusters were rejected because of this cut, leaving us with a final sample of 75 clusters with complete data available (see Table 5 of Collins et al. 1994). The position angles of these clusters formed the basis for our search for statistically significant cluster alignments and will be published soon in our forthcoming data paper, Collins et al. (1994).

### 3.2 Errors on the Fitted Position Angles

For any statistical study of cluster alignments, it was vital that we quantified the error on our fitted ellipse since this is the major limit on the accuracy with which we can study cluster alignments. An obvious check of this would be to compare our position angles with those presented in the literature (e.g Figure 1). Such a comparison is hampered by the fact that our clusters are located in the Southern Hemisphere, while previous cluster alignment studies have been concentrated in the Northern Hemisphere. In total, we were only able to find one cluster in common between our sample and those of West (1989b) and Plionis (1994); Abell cluster A14 for which West quotes a position angle of $53°$ compared to our computed position angle of $35°$ (see Collins et al. 1994). For this single case, we are well within the typical errors shown in Figure 1 for the aforementioned works and discussed below for our sample. A further simple consistency check we made was to compare the cumulative distribution of our position angles with that of a uniform distribution, see Figure 2. A Kolmogorov–Smirnov (K–S) test confirms that the two distributions are probably drawn from the same parent distribution, indicating that we have no systematic bias toward a particular position angle in our sample.

In the spirit of other work in this field, we carried out Monte–Carlo simulations of our clusters to estimate the standard error on our fitting procedure given above. In an attempt to simulate our fitting procedure as closely as possible, we took all clusters in our sample discussed above as the basis for our simulations. We therefore, constructed 75 model clusters with the same characteristics (richness, back-ground *etc*) as our real dataset. The procedure we used for each individual cluster is described here.

(i) The radial profile of the real cluster was fitted with a Gaussian.
(ii) We then determined the number–magnitude relationship for the central $1.0h^{-1}$ Mpc region of our 1 deg$^2$ area taken from the EDSGC which is centred on the real cluster. We also determined the number–magnitude relationship for the surrounding background and subtracted this, after it had been normalised by their respective areas, from the cluster number–magnitude relationship.
(iii) We took the profile of our real cluster discussed above and constructed a simulated cluster with an ellipse of known position angle and an eccentricity equal to the mean eccentricity observed from our sample.
(iv) Using the background corrected number–magnitude counts for the cluster, we randomly scattered the same number of galaxies within the boundary of the simulated ellipse (above) ensuring that the galaxies had the same Gaussian profile as the real cluster. A random background to the cluster was constructed using the observed background number–magnitude relationship for that cluster discussed above.
(v) Therefore, this provided us with a simulated cluster that had the same general profile, the same magnitude distribution and the same background as the original cluster. The only free parameter of the simulated cluster was the position angle.

For each cluster, we carried out 20 simulations randomly changing the position angle of each model cluster between each of the simulations. This allowed us to determine the standard deviation of the difference between the final fitted position angles of the model clusters and the actual position angles set at the start of the fitting procedure. Finally, we took the average of these standard deviations which we quote as the mean uncertainty on our fitting procedure and is presented as an error bar on Figure 4 ($\simeq \pm 10°$). In addition, Figure 5 presents the individual standard deviations (described as error on $\theta$) for each of our model clusters as a function of their fitted richness.

The advantage of this method is that it mimics our observed clusters as closely as possible. Therefore, this method should be a fair representation of the error associated with fitting an ellipse to our clusters using only a finite number of galaxy positions.

## 4  CLUSTER ALIGNMENTS

In the original work by Binggeli (1982), he observed a general trend that clusters of galaxies appeared to point towards their nearest neighbour up to a distance of $30h^{-1}$ Mpc. In addition, he noted that there was a tendency for clusters to point towards each other over a considerably larger distance of $50 - 100h^{-1}$ Mpc. These observations highlight the two types of cluster alignment we are considering here. First, to determine if a cluster is elongated, in projection, towards its closest spatial neighbour, irrespective of the orientation of that cluster. Secondly, to establish if there is a general tendency for clusters to point towards each other and discover if there is a coherent orientation between clusters within the same supercluster or between superclusters. We should mention here that we are dealing with the projected orientation of the clusters, not their true spatial orientation. However, for the majority of cases the two will certainly be related and consistent with each other.

In this section, we present the spherical trigonometry required to derive the alignment between two clusters. We do this for completeness because we could not find an adequate discussion of the trigonometry in any previous paper on this subject.

We define the pointing angle ($\theta_{PA}$) as the angle between a cluster's major axis and the great circle connecting it and another cluster. Figure 3 is a schematic diagram which illustrates the problem of determining the pointing angle for a particular cluster. In this diagram, the two clusters in question are at the vertices $b$ and $c$ of the spherical triangle $abc$, where $a$ is the pole of the celestial sphere (either north or south). The sides $\widehat{ac}$ and $\widehat{ab}$ are lines of constant Right Ascension to each of the clusters, while the third side $\widehat{bc}$ is the great circle joining the two clusters. For this spherical triangle we known the length of all three sides and one of the angles, which are:

$$\begin{aligned}
\theta_a &= \Delta\alpha, \\
\widehat{bc} &= \theta_{sep}, \\
\widehat{ac} &= \frac{\pi}{2} - \delta_1, \\
\widehat{ab} &= \frac{\pi}{2} - \delta_2, \quad (5)
\end{aligned}$$

where $\Delta\alpha$ is the difference in the Right Ascension of the clusters and $\delta_1$, $\delta_2$ are their Declinations. The angular separation between the clusters, $\theta_{sep}$, is given by the standard formula,

$$\cos\theta_{sep} = \sin\delta_1 \sin\delta_2 + \cos\delta_1 \cos\delta_2 \cos(\alpha_2 - \alpha_1). \quad (6)$$

Using standard cosine rules for spherical triangles, we were able to determine the two remaining angles of this spherical triangle.

The pointing angles of the two clusters can be determined from the two angles of the spherical triangle $\theta_b$ & $\theta_c$ and the measured position angles of the clusters (as shown in the section 3.1, the position angle is the clockwise angle of the major axis away from the line joining the cluster with the pole *i.e.* angle $\theta$ in Figure 3.) The exact method of deriving the pointing angle of each cluster depends on the respective positions and quadrant's of the two clusters *i.e.* whether the accompanying cluster is east or west of the cluster in question. For example, in Figure 3, the cluster at $c$ has a pointing angle defined by $\theta - \theta_b$, while for the cluster at $b$, the pointing angle is $180 - \theta - \theta_c$.

## 5  RESULTS

To investigate the first of the two alignment effects discussed in the previous section, we have plotted the pointing angle towards a clusters' nearest neighbour against the spatial separation of the two clusters (Figure 4). This is a replication for our data of the original plot made by Binggeli (1982). This plot does show some evidence for an alignment effect between nearest neighbouring clusters at small separations ($< 10h^{-1}$ Mpc), since there appears to be an excess of pairs with low pointing angle values. For example, the mean pointing angle for the 9 clusters with a nearest neighbour separation of $\leq 5h^{-1}$ Mpc is $17° \pm 13°$. However, the mean pointing angle increases to $43° \pm 30°$ for the next 9 pairs in the separation range $5 < r \leq 10h^{-1}$ Mpc, which is consistent with that expected from a random distribution ($\langle\theta\rangle \simeq 45°$) Clearly, we are not seeing an alignment effect on the scale advocated by Binggeli (1982).

Also plotted in this figure, is the error obtained from our Monte–Carlo simulations discussed above, which we observed to be $\sim \pm 10$ degrees and was mostly insensitive to the eccentricity, richness and redshift of the clusters. However, as can be seen in Figure 5, which shows the measured error (the standard deviation between the fitted position angle and imposed actual position angle for 20 model clusters) versus richness for each of our individually simulated clusters. This figure shows that the error was only stable for clusters with $> 40$ galaxies in the final fitted ellipse. Below this, the uncertainty on the position angle rises dramatically.

As a further check of our fitted parameters, Figure 6 shows the number of fitted galaxies in the final ellipse against redshift of the cluster. This plot indicates two important points: *(a)* The majority of our cluster have a fitted richness greater than $n = 40$, which is reassuring in the light of our simulations; *(b)* There is no significant correlation which strongly suggests that we are not preferentially selecting richer clusters at higher redshift.

| Sample | $\langle\theta\rangle$ | $\sigma/\sqrt{N}$ |
|---|---|---|
| $r < 10h^{-1}$ Mpc | 32.8 | 4.9 |
| $r < 10h^{-1}$ Mpc Random | 40.5 | 4.4 |
| $10 \leq r < 30h^{-1}$ Mpc | 48.4 | 3.7 |
| $10 \leq r < 30h^{-1}$ Mpc Random | 45.8 | 3.6 |
| $30 \leq r < 60h^{-1}$ Mpc | 47.7 | 1.4 |
| $30 \leq r < 60h^{-1}$ Mpc Random | 45.0 | 1.2 |
| $60 \leq r < 100h^{-1}$ Mpc | 45.5 | 1.0 |
| $60 \leq r < 100h^{-1}$ Mpc Random | 46.3 | 1.0 |

**Table 1.** The mean pointing angle for all pairs in different samples as a function of spatial separation. The standard error is quoted on all the numbers. The samples marked "Random" are the same as the real data samples but with their position angles drawn from a random uniform distribution. In other words, this would be the expected result for no alignment effect given the number of cluster pairs involved in the calculation.

To statistical study the second form of cluster alignments mentioned above, that clusters tend to point towards each other up to separations of $100h^{-1}$ Mpc, we employed both the standard methods used in the literature (West 1989b) and a Spatial Two–Point Correlation Function analysis. Table 1 shows the mean pointing angle, as a function of separation, for all pairs in our sample of clusters described earlier. For the case of no alignment effect, the mean should be $\simeq 45°$, while a significant alignment would skew the mean to a lower value. For the smallest separations ($r < 10h^{-1}$ Mpc), there is some evidence that this may be the case. However, the standard error on this measurement is large enough to make it consistent with the expected value *i.e.* $< 3\sigma$. Other samples presented in Table 1 at larger spatial separations are fully consistent with no alignments. This is highlighted by the fact that our observed mean pointing angles are indistinguishable from those derived from datasets with randomised position angles.

In addition to the computed mean pointing angles above, we carried out a K–S test on the distribution of pointing angles in our sample, in different spatial separation intervals. The results of this test are shown in Figure 7, where we have plotted the cumulative distribution of our pointing angles compared to that expected for a uniform distribution *i.e.* no alignment effect. Again, these results present tantalising evidence for alignments at the smallest separations, in agreement with both the mean pointing angle analysis and the Binggeli plot. Once more, however, the statistical significance of this observation is not compelling, since the probability that the null hypothesis (that they are drawn from the same distribution) is still acceptable at the 7% level. All other K–S tests, for larger separations, are fully consistent with a uniform distribution.

Finally, we implemented a Spatial Two–Point Correlation Function analysis (Peebles 1980), which we define as;

$$\xi_{cc}(r, \theta_{PA}) = \frac{2 N_{ran}}{N_c} \frac{n_{cc}(r, \theta_{PA})}{n_{cr}(r, \theta_{PA})} - 1, \qquad (7)$$

where $n_{cc}$ is the binned cluster–cluster pairs as a function of both pointing angle ($\theta_{PA}$) and spatial separation (r), while $n_{cr}$ is the binned cluster–random pairs. For this task, we generated a random catalogue of clusters 100 times greater than the data catalogue ($N_{ran} = 100 N_c$) covering the exact same area as the data as well as having the same redshift distribution, after it had been smoothed with a Gaussian of width 3000 km s$^{-1}$. The position angles of the random clusters were selected from a uniform distribution.

The advantage of this analysis is it allows us to study the full relationship between pointing angle and spatial separation, without constraining ourselves to selected intervals in separation as above. One disadvantage is that it involves binning the data, which may swamp a small signal if present. Since this is the first time the correlation function has been used to study such alignment effects, we carried out a simulation to determine the sensitivity of this statistic to cluster alignments. We artificially inserted the *Binggeli Effect* into our real data by assigning each cluster a position angle so it would point, within $\pm 10$ degrees (the error quoted above from our simulations), towards its nearest neighbour ($r \leq 30h^{-1}$ Mpc). The result of this simulation is shown in Figure 8 which clearly shows that the correlation function has detected the effect. As expected, the function declines rapidly as both a function of spatial separation and pointing angle, with the contours becoming negative beyond $30h^{-1}$ Mpc. Figure 9 shows a slice of the correlation function taken along the x–axis (constraining spatial separations to $< 10h^{-1}$ Mpc), which demonstrates the high significance to which the artificially imposed alignment effect can be seen (the error bars are possionian derived from $\delta\xi = (1 + \xi)/\sqrt{n_{cc}}$). The above simulations were repeated with the imposed error on the artificial pointing angles between neighbouring clusters increased to $\pm 30$ degrees (reflects the scatter seen in Figure 1). The spatial correlation function once again securely detected the presence of the effect, but at a lower significance level.

Figures 10 shows the result of our correlation analysis for all cluster pairs in our sample irrespective of nearest neighbours. This contour plot does exhibit a positive peak at the origin of our $\xi_{cc}(r,\theta_{PA})$, which is consistent with an alignment effect as demonstrated by our simulations above. The measured $\xi_{cc}(r,\theta_{PA})$ remains positive out to separations of $\simeq 20h^{-1}$ Mpc. However, it should be noted that the lowest positive contour, as plotted in Figure 10, does extend over a large range of pointing angles (x–axis of the figure) between $0°$ and $\simeq 60°$. This is highlighted in Figure 11, where we present a slice along the x–axis of Figure 10 and which shows that $\xi_{cc}(r < 10, \theta_{PA})$ remains positive across almost the entire possible range of pointing angles. For a strong alignment effect, the simulation show us that the contours should fall isotopically from the origin, which is not the case here. Figure 11 also presents the possionian error bars (defined above) on our measurement of the correlation function. The statistical significance of these positive contour values is not conclusive, which is consistent with previous statistics used on this dataset.

Finally, to test the claim that cluster alignments are the strongest within the same host supercluster, we confined our correlation analysis to a sample of clusters selected to be within superclusters. Guzzo et al. (1992) has shown that the central two hours of Right Ascension of the EM survey ($23^{hrs} \rightarrow 1^{hrs}$) is dominated by two of the largest superclusters known; the Sculphor Supercluster at a redshift of 0.11, centered on the SGP, has $\sim 15$ rich clusters. Therefore, by constraining our analysis to this central region, we should be dominated, at separations $r < 50h^{-1}$ Mpc, by cluster pairs from within the same host supercluster.

The results of this analysis are very similar to that derived above for the whole sample of clusters. We found tentative evidence for small scale alignments ($r < 10h^{-1}$ Mpc) but again they were not statistically significant. On larger scales, the correlation function is consistent with no alignment effect. However, this analysis only involved 46 clusters and therefore, we should be careful not to overinterpret the significance of this result.

# 6 DISCUSSION

From our analysis, we see some evidence for cluster alignments on spatial scales of less than $10h^{-1}$ Mpc, however, the effect is not statistically significant. On larger scales, we see no evidence at all, statistically significant or not, for cluster alignments. Our result is in most disagreement with the work of West (1989b) and Plionis (1994) who claim to find strong alignments between clusters up to separations of $\sim 60h^{-1}$ Mpc.

This observational discrepancy between our results may be due to the relative sizes of our respective samples of clusters and it could be argued that we suffer from small number statistics, since we only have 75 clusters. However, if this was the case, it is hard to understand why we see a possible alignment effect at small separations where we have the least number of cluster pairs. For example, in Figure 7 the number of cluster pairs in the smallest spatial interval ($r < 10h^{-1}$ Mpc) is 52 compared to the $30 < r < 60h^{-1}$ Mpc interval where we have 720 pairs and see no effect. To investigate this point further, we implemented two cluster alignment searches with increased numbers of cluster using the EDCC. For clusters without a measured redshift we estimated their redshift using the best determination of the $\log z - m_{10}$ relationship (Nichol 1992; error of $\simeq 20\%$). The first search consisted of 126 EDCC clusters with richness $> 40$, while the second involved all EDCC overdensities for which we could determine the position angle i.e. iterated to a stable solution. This gave us a sample of 527 overdensities. The results of this analysis are almost identical to those presented above for our sample of 75 clusters, which is somewhat surprising considering the likely large scatter introduced by the estimated redshifts. In light of this, we feel our small initial sample size, with respect to previous studies, does not play a crucial role, especially at the larger separations. At separations of $r < 10h^{-1}$ Mpc, a larger sample of automated clusters would certainly increase the statistical significance of the alignment effect we tentatively observe.

The major difference between the analysis presented here and those of West (1989b) and Plionis (1994) is in the methods used to select our respective cluster samples. We therefore, compare our findings with other published work which utilise's non–Abell catalogues of clusters/groups. Our findings agree, in terms of spatial scale and statistical significance, with those published by Fong et al. (1990), who also analysed an objective sample of overdensities. Their sample was selected from plate scans of photographic plates like ours, but their overall survey area is significantly smaller than that presented here. Furthermore, their survey area was not contiguous thus limiting their analysis to scales of $r < 30h^{-1}$ Mpc. West (1989a) carried out an analysis on groups of galaxies objectively selected from the CfA and SRSS redshift surveys. West found a lower level of alignment than that found for clusters of galaxies ($r < 20h^{-1}$ Mpc), which is more consistent with our work. However, the alignment effect is *only* seen in the SRSS survey with no alignment effect on any scale detected in the CfA survey. West attributes this major discrepancy to a higher interloper contamination in the Northern survey. This is easy to understand since the selected groups are intrinsically poor containing only a few galaxies each, thus one interloper could significantly effect the measured position angle. Clearly, this and our simulations, demonstrates the dangers of working with such poor systems. Finally, Ulmer et al. (1989) have used a heterogeneous sample of 45 bright X–ray clusters to search for the existence of cluster alignments. They found no statistical evidence for strong cluster alignments which is consistent with us. In addition, for clusters with both optical and X–ray determinations of the position angle, they showed that at least 50 galaxies were required to accurately estimate the orientation of the cluster. This is again in agreement with the findings presented here. Therefore, it would appear that many of the objectively constructed samples of overdensities do not exhibit the same level of alignments as seen in the larger Abell/Lick samples.

The original motivation behind our work was to place this area of study on a stronger observational footing. For the first time, we have searched for cluster alignments in a statistically complete, objective sample of clusters. However, as a final comment we should note the limitations of our analysis here. Both Figure 1 and our own simulations highlight the difficulty involved in estimating the position angle of clusters from a finite number of galaxies. Clearly, this will always be a problem for optical studies. Furthermore, we have taken no account of possible subclustering in our clusters which could have a significant effect on any optically derived result. An obvious direction to pursue in this field is the use of X–ray data, since the hot X–ray emitting gas delineates the cluster potential well directly as well as allowing any subclustering to be quantified. Such analysis may be possible with the advent of objectively constructed, statistical samples of X–ray clusters from the ROSAT All–Sky survey.

## 7 ACKNOWLEDGEMENTS

We graciously thank Mel Ulmer for both his many stimulating discussions and informative advice during this project. David Martin also thanks him for providing financial support during this work. We are especially grateful to Mike West, our referee, for his refreshing openness and co–operation during the refereeing process of this paper. His help and input certainly made this paper better. Dave Martin thanks Jim Peebles for helpful comments and Bob Nichol thanks Avery Meiksin for useful suggestions. This work was supported in part by a NASA Space Consortium Grant through Aerospace Illinois.

## 9 FIGURE CAPTIONS

**Figure 1:** A comparison of 34 Abell clusters that have a robust position angle determined independently by both Plionis (1994) and West (1989b). The error bars plotted here were also determined independently by both authors. The dotted line represents the expected one–to–one relationship. The mean difference, and scatter, between the two sets of measurements is given as well.

**Figure 2:** The normalised cumulative distribution for the position angles presented in this paper. The dashed like represents a uniform distribution. The K–S probability that these two distributions are drawn from the same parent distribution is also given.

**Figure 3:** The number of galaxies within the fitted ellipse against the observed redshift of the cluster.

**Figure 4:** A replica of the plot shown by Binggeli (1982) using clusters from the EM survey. This shows the angle between the great circle connecting a cluster to its nearest neighbouring cluster and the major axis of that cluster (pointing angle $\theta_{PA}$) against their spatial separation (r). The bar represents our estimation of the error on determining the cluster pointing angles (see text for explanation).

**Figure 5:** The error in radians on the fitted ellipse for our individual simulated clusters (the standard deviation between the actual imposed position angle and the eventual fitted position angle for 20 simulations) versus the richness of those simulated clusters.

**Figure 6:** Schematic diagram of the spherical trigonometry involved in computing the alignment between two clusters on a celestial sphere.

**Figure 7:** The normalised cumulative distribution of pointing angles for our clusters in different spatial separation intervals. The dashed lines represented the expected distributions for a uniform sample *i.e.* in the absence of alignment effects. The K–S test probability that these two distributions are draw from the same parent distribution is also shown.

**Figure 8:** The Spatial Correlation Function for our cluster dataset with an artificial Binggeli alignment effect introduced *i.e.* clusters point to their nearest neighbour within $\pm 10$ degrees (our simulated error given in the text) over a distance of $30h^{-1}$ Mpc. The x–axis is spatial separation of cluster pairs, while the y–axis is the pointing angle of the clusters. The contour levels are: -0.9, -0.6, -0.3, 0.1, 0.6, 1.3, 2.1, 3.1, 4.4, 6.0. Clearly, the correlation function has detected the exact magnitude of our induced cluster alignment.

**Figure 9:** The same data set as used in Figure 8, but this time the spatial coordinate in Figure 8 has been constrained to $r \leq 10h^{-1}$ Mpc. This provides a clearer display of the effect along with the error bars (see text).

**Figure 10:** The Spatial Correlation Function for our real data described in the text. This plot shows the observed corrections of the pointing angle between clusters. The contour levels are: -0.9, -0.7, -0.5, -0.3, 0.0, 0.3, 0.7, 1.1, 1.6, 2.2. There is tentative evidence for an alignment effect out to a cluster separation of $20h^{-1}$ Mpc, but as shown in Figure 10, it is not statistically compelling.

**Figure 11:** The same Spatial Correlation Function as presented in Figure 9, but with the spatial coordinate constrained to r $\leq 10h^{-1}$ Mpc. This shows the magnitude, and error bars, of the positive correlations see along the x–axis in Figure 10.